\documentclass[12pt]{iopart}
\usepackage{graphicx}
\usepackage{dcolumn}
\usepackage{array}
\usepackage{iopams} 
\usepackage{bm}

\begin{document}

\title[Auxiliary fields and analytical solutions of the Schr\"{o}dinger equation]
{Auxiliary fields as a tool for computing analytical solutions of the
Schr\"{o}dinger equation}

\author{Bernard Silvestre-Brac$^1$, Claude Semay$^2$ and Fabien Buisseret$^2$}

\address{$^1$ LPSC Universit\'{e} Joseph Fourier, Grenoble 1,
CNRS/IN2P3, Institut Polytechnique de Grenoble, 
Avenue des Martyrs 53, F-38026 Grenoble-Cedex, France}
\address{$^2$ Groupe de Physique Nucl\'{e}aire Th\'{e}orique, Universit\'{e}
de Mons-Hainaut, Acad\'{e}mie universitaire Wallonie-Bruxelles, Place du Parc 20,
B-7000 Mons, Belgium}
\eads{\mailto{silvestre@lpsc.in2p3.fr}, \mailto{claude.semay@umh.ac.be}, 
\mailto{fabien.buisseret@umh.ac.be}} 

\date{\today}

\begin{abstract}
We propose a new method to obtain approximate solutions for the Schr\"{o}dinger equation with an arbitrary potential that possesses bound states. This method, relying on the auxiliary field technique, allows in many cases to find analytical solutions. It offers a convenient way to study the qualitative features of the energy spectrum of bound states in any potential. In particular, we illustrate our method by solving the case of central potentials with power-law form and with logarithmic form. For these types of potentials, we propose very accurate analytical energy formulae which improve a lot the corresponding formulae that can be found in literature.  
\end{abstract}

\pacs{03.65.Ge}
\maketitle

\section{Introduction}

From the early years of quantum mechanics, searching for analytical solutions of the Schr\"{o}dinger equation has always been a subject of intense investigations, in particular when the Hamiltonian possesses bound states. This is a challenging mathematical problem often involving complex differential equations, and several methods have been proposed to find approximate analytical solutions: WKB method, semiclassical treatment, variational method, perturbation theory, etc. More details can be found in~\cite{flu} for example, in which a wide range of analytical results concerning bound states are presented. Nowadays, numerical calculations allow us to solve eigenequations very accurately. But, finding analytical solutions is still of great interest since it makes explicitly appear the dependence of the eigenenergies and wave functions on the various parameters of the model. Such a property is particularly useful when one tries to fit the parameters of a model to experimental data for example. In the present work, we propose a new way to obtain analytical approximate solutions of Schr\"{o}dinger's equations for bound state problems. Our method is based on the auxiliary field technique. 

Auxiliary fields, also known as einbein fields, appear in various domains of theoretical physics. They are commonly used to get rid of the square roots appearing in relativistic systems; let us mention the Nambu-Goto Lagrangian which is replaced by the Polyakov one in string theory for example~\cite{str}. They also appear in supersymmetric field theories~\cite{susy} and in hadronic physics~\cite{af1}. Let us briefly present the auxiliary fields with the simple example of a free relativistic particle, that is the Lagrangian
\begin{equation}
\label{exa1}
{\cal L}=-m\, \sqrt{\dot{\rm x}^2},
\end{equation}
where $\dot{\rm x}$ is the world-velocity.
An auxiliary field $\mu$ can be introduced in this expression to get rid of the square root
\begin{equation}
\label{exa2}
{\cal L}(\mu)= \frac{\dot{\rm x}^2}{2\mu}+\frac{m^2}{2}\mu.
\end{equation}
This last expression is formally simpler than Lagrangian~(\ref{exa1}), but it is equivalent if the equations of motion of the auxiliary field are considered. The auxiliary field is determined by the condition of extremum
\begin{equation}
\left.\delta{\cal L}(\mu)\right|_{\mu=\mu_0}=0\Rightarrow \mu_0=-\frac{\sqrt{\dot{\rm x}^2}}{m},
\end{equation}
and if we substitute it into the corresponding Lagrangian (\ref{exa2}), we recover the original one (\ref{exa1}): ${\cal L}(\mu_0)={\cal L}$. This is a general feature of auxiliary fields: Their equation of motion are not dynamical but rather algebraic relations that express them in terms of the other variables of the problem. 

We see that the introduction of auxiliary fields can notably simplify the formalism, and thus the computations. It was pointed out in~\cite{lucha2} that relativistic Hamiltonians can be rewritten as apparently nonrelativistic ones by such a method [compare equations~(\ref{exa1}) and (\ref{exa2}) for example]. We previously showed that auxiliary fields allow to express a linear potential, which is widely used in hadronic physics, as an expression involving an auxiliary field and a harmonic oscillator~\cite{af3}. By doing this, analytical mass spectra can be computed assuming that the auxiliary field is not an operator but a real number. That is why approximate results are obtained. The question is: Could any arbitrary potential be expressed in terms of an equivalent expression involving an auxiliary field and another potential for which an analytical solution is known? As we will show in section~\ref{grmeth}, the answer is yes, and a general algorithm for analytically solving the Schr\"{o}dinger equation with any potential can be proposed. In sections~\ref{hoapp} and \ref{capp}, we illustrate our general method by computing analytical energy spectra for central potentials with power-law form and with logarithmic form. As our formulae are approximate, it is of interest to try to maximally improve them. This is done in section~\ref{impfor}. Finally, we draw some conclusions in section~\ref{conclu}.

\section{General method}
\label{grmeth}

We begin with a standard Hamiltonian containing a kinetic energy term $T(\bm p^{\, 2})$ and a local central potential $V(r)$ 
\begin{equation}
\label{inco}
H=T(\bm p^{\, 2})+V(r),
\end{equation}
whose eigenvalues and eigenstates are a priori not analytically known. On the contrary, let us assume that the Hamiltonian with potential $P(r)$ replacing $V(r)$, namely $H_A(\rho)=T(\bm p^{\, 2})+\rho\, P(r)$, where $\rho$ is a real parameter, possesses bound states with an analytical spectrum. Thus, we explicitly know the solution of the corresponding Schr\"{o}dinger equation
\begin{equation}
\label{analy}
H_A(\rho) \left|\Psi(\rho)\right\rangle=E_A(\rho)\left|\Psi(\rho)\right\rangle.
\end{equation}
Starting from the eigenequation~(\ref{analy}), we propose hereafter a way to compute an analytical approximate solution of the eigenequation
\begin{equation}
H \left|\Psi\right\rangle=E\left|\Psi\right\rangle,	
\end{equation}
$H$ being the arbitrary Hamiltonian~(\ref{inco}). The procedure consists in three steps:
\begin{enumerate}
\item Build the function $K(r)$ and the operator $\hat\rho$ by the definitions
\begin{equation}
\label{rdef}
\hat\rho=\frac{\partial V(r)/\partial r}{\partial P(r)/\partial r} \equiv K(r),
\end{equation}	
so that the operator $r$ is given by the inverse function
\begin{equation}
\label{invfuncKk}
r = K^{-1}(\hat\rho).
\end{equation}
\item Introduce the Hamiltonian $\tilde H(\rho)$ through the definition  
\begin{equation}
\label{htdef}
\tilde H(\rho)=T(\bm p^{\, 2})+\rho\, P(r)+V\left[K^{-1}(\rho)\right]-\rho\,
P\left[K^{-1}(\rho)\right].
\end{equation}
$\rho$ is known as an auxiliary field.
Because of~(\ref{analy}), the spectrum of $\tilde H(\rho)$ is analytically known if the auxiliary field is assumed to be a number, and not an operator. The eigenstates are $\left|\Psi(\rho)\right\rangle$, while the eigenenergies $E(\rho)$ are obtained from $E_A(\rho)$ simply by adding a $\rho$-dependent constant term
\begin{equation}
\label{en1}
E(\rho)=E_A(\rho)+V\left[K^{-1}(\rho)\right]-\rho\, P\left[K^{-1}(\rho)\right].
\end{equation}
\item The physical spectrum is finally given by the eigenstates $\left|\Psi(\rho_0)\right\rangle$ and the eigenenergies $E(\rho_0)$, with $\rho_0$ minimizing the total energy~(\ref{en1}), i.e.
\begin{equation}
\label{enmin}
\left.\frac{\partial E(\rho)}{\partial \rho}\right|_{\rho=\rho_0}=0.
\end{equation}
\end{enumerate}

It is now time to justify the procedure given above. The Hamiltonian $H$ is replaced by $\tilde H(\rho)$, given by~(\ref{htdef}). Actually,
one replaces in $H$ the potential $V(r)$, for which the energies are not analytical, by the potential $P(r)$ for which they are, but via an effective potential $\tilde V(r,\rho)=\rho\, P(r)+V\left[ K^{-1}(\rho)\right]-\rho\, P\left[K^{-1}(\rho)\right]$ depending on the auxiliary field.
Such an auxiliary field is nothing else but a mathematical trick to simplify the formalism. In the present case for example, $V( r)$ is ``simulated" by a function of $\rho$ and of an analytically solvable potential $P(r)$. The key point is that the auxiliary field, when properly eliminated as an operator, should lead to Hamiltonian~(\ref{inco}). It can be checked that 
\begin{equation}
\left.\delta \tilde V(r,\rho)\right|_{\rho=\hat\rho}=0\Rightarrow \hat\rho=K(r).
\end{equation}
Moreover, $\tilde V(r,K(r))=V(r)$. This means that Hamiltonians~(\ref{inco}) and (\ref{htdef}) are actually equivalent up to the elimination of the auxiliary field as an operator, the auxiliary field being then defined by~(\ref{rdef}).   

The approximation in our method is the following: If $\rho$ is considered as a number in Hamiltonian~(\ref{htdef}), the corresponding eigenequation can be analytically solved, since only $P(r)$ is involved. Then, an energy spectrum of the form~(\ref{en1}), which depends on the auxiliary field, is obtained. Obviously, the physical value of $\rho$ for a given eigenenergy is the one which ensures this energy to be minimal. This is expressed in~(\ref{enmin}). 

The Hellmann-Feynman theorem \cite{feyn} states that 
\begin{equation}
\frac{\partial E(\rho)}{\partial \rho}=
\left\langle \Psi(\rho)\right| \frac{\partial \tilde H(\rho)}{\partial \rho} \left|\Psi(\rho)\right\rangle.
\end{equation}
Using this relation and defining the quantity $r_0$ by $\rho_0 = K(r_0)$, 
one can show that
\begin{equation}
\label{pr0}
\left\langle \Psi(\rho_0)\right| P(r) \left|\Psi(\rho_0)\right\rangle = P(r_0).
\end{equation}
This means that $r_0$ is a kind of ``average point" for the potential $P(r)$. If functions $P(r)$ and $V(r)$ are smooth enough, one could expect that the average point for the function $K(r)$ does not differ significantly from $r_0$. So we can assume that
\begin{equation}
\left\langle \Psi(\rho_0)\right| \hat \rho \left|\Psi(\rho_0)\right\rangle \approx 
K(r_0)=\rho_0.
\end{equation}
Thus, we can expect that the optimal value $\rho_0$ should be close to the average value $\left\langle \Psi(\rho_0)\right|\hat\rho\left|\Psi(\rho_0)\right\rangle$, with $\hat \rho$ given by~(\ref{rdef}) \cite{sema04}. Our method could actually be considered as a ``mean field approximation" with respect to a particular auxiliary field which is introduced to simplify the calculations. This will be checked in the following. 

It is possible to obtain an analytical estimation of the accuracy of the method. With the notations introduced above, let us compute $\langle \Psi(\rho_0)| \tilde H(\rho_0)-H|\Psi(\rho_0)\rangle$. Using (\ref{inco}), (\ref{htdef}) and (\ref{pr0}), we find
\begin{equation}
\label{errana}
E(\rho_0)-\langle \Psi(\rho_0)| H |\Psi(\rho_0)\rangle = V(r_0)-\langle \Psi(\rho_0)| V(r) |\Psi(\rho_0)\rangle.
\end{equation}
The r.h.s.\ of this equation is the difference between the value of potential $V$ computed at the ``average point" $r_0$ and the average of this potential for the trial state $ |\Psi(\rho_0)\rangle$. In some favorable cases \cite{macd33} (the trial state is a ground state for instance), $E \le \langle \Psi(\rho_0)| H |\Psi(\rho_0)\rangle$, where $E$ is the exact eigenvalue of $H$. Since this inequality cannot be always guaranteed, we just can write in general
\begin{equation}
\label{errana2}
E(\rho_0)-E \gtrsim V(r_0)-\langle \Psi(\rho_0)| V(r) |\Psi(\rho_0)\rangle.
\end{equation}
In principle, all the quantities present in this formula, except $E$ obviously, can be analytically computed. So an estimation of the error on the absolute eigenvalue can be obtained. Depending on the sign of the r.h.s.\ of (\ref{errana2}), $E(\rho_0)$ can be an upper or a lower bound of $E$. An estimation of the relative error is obtained by computing $(V(r_0)-\langle \Psi(\rho_0)| V(r) |\Psi(\rho_0)\rangle)/|E(\rho_0)|$. Numerical evaluations of formula~(\ref{errana2}) will be given in the following. 

Formally, we thus have proposed an algorithm which allows to compute approximate analytical eigenvalues and eigenstates of an arbitrary Hamiltonian containing the potential $V(r)$, provided that an analytical solution of the Hamiltonian is known for a particular potential $P(r)$. This method leads to approximate solutions because of the presence of the auxiliary field, but, as we will see, it is a very convenient tool to understand the qualitative features of various energy spectra.

Before giving explicit examples, three points should be stressed:
\begin{itemize}
\item One easily checks that the method is exact for $V(r)=P(r)$, with $\hat \rho=1$ trivially. One thus expects that, the closer is $P(r)$ to $V(r)$, the better the approximation is.
\item Generally, the procedure is not variational, so that we have no certainties concerning the position of the approximate energies with respect to the exact ones.
\item The method is not free of technical difficulties. Although the presented algorithm is completely general and in principle applicable to any situation, obtaining an analytical solution is not always possible or can lead to cumbersome calculations. In fact the crucial difficulty concerns the inversion of the function $K(r)$ [formula~(\ref{invfuncKk})] and/or the determination of the best auxiliary field $\rho_0$ [formula~(\ref{enmin})]. However, these difficulties can be overcome in many relevant cases and the corresponding equations can be fortunately solved. 
\end{itemize}
We can also notice that the method which is proposed here remains valid not only for central potentials, but also for general quantum mechanical systems (one- and two-dimensional ones for example).

\section{The case $P(r)=r^2$}
\label{hoapp}

\subsection{Power-law potentials}

Let us first illustrate the general procedure presented in the previous section by solving the eigenequation associated to the nonrelativistic Hamiltonian
\begin{equation}
\label{hzero}
H=\frac{\bm p^{\, 2}}{2m}+\textrm{sgn}(\lambda)\, a\, r^\lambda,
\end{equation}
with $\lambda \ne 0$, $\textrm{sgn}(\lambda)=|\lambda|/\lambda$, and $a>0$ in order that this Hamiltonian possesses bound states. 

Two power-law potentials are standard in quantum mechanics formalism: The harmonic and Coulomb potentials because analytical solutions are known whatever the radial and orbital quantum numbers. This should thus be of interest to take $P(r)=r^2$ or $-1/r$ in our auxiliary field method. In this section, we will consider the case of the harmonic oscillator. We recall that the spectrum of the Schr\"{o}dinger equation with potential
\begin{equation}
\label{ehp}
V^H(r)=\frac{1}{2}m\omega^2r^2
\end{equation}
is given by (see for example~\cite[problem 66]{flu})
\begin{equation}\label{nrho1}
R^H(r)= N_{n\ell}\ r^\ell\ e^{-m\omega r^2/2}\ L^{\ell+1/2}_n(
m\omega\, r^2), \quad  E^H=\omega(2n+\ell+3/2),
\end{equation} 
where $N_{n\ell}$ is a normalization coefficient and $R^H(r)$ the radial wave function ($n=0,1,2,\ldots$ is the radial quantum number and $\ell=0,1,2,\ldots$ is the orbital quantum number). 

Taking $P(r)=r^2$,~(\ref{rdef}) leads to 
\begin{equation}
\label{rdpl}
\hat\rho=\frac{a|\lambda|}{2}r^{\lambda-2}\Rightarrow r = K^{-1}(\hat\rho)=\left(\frac{2\hat\rho}{a|\lambda|}\right)^{\frac{1}{\lambda-2}}.
\end{equation}
We notice that for $\lambda \neq 2$, $K^{-1}(\hat\rho)$ is always well-defined since $a|\lambda| >0$ by definition. The case $\lambda=2$ is actually of no interest since in this case $\hat\rho=1$ trivially, and the results are exactly known. Then, following~(\ref{htdef}), one obtains after some algebra
\begin{equation}
\label{hpl1}
\tilde H(\rho)=\frac{\bm p^{\, 2}}{2m}+\rho\, r^2 + \frac{2-\lambda}{2\lambda} 
(2 \rho)^{\frac{\lambda}{\lambda-2}}(a|\lambda|)^{\frac{2}{2-\lambda}}.
\end{equation}
We can easily check that Hamiltonian~(\ref{hpl1}) is fully equivalent to Hamiltonian~(\ref{hzero}) if $\rho=\hat\rho$ as given by~(\ref{rdpl}). In order to go further in the computation, we now consider the auxiliary field $\rho$ as a real parameter. As we already mentioned, this is at this place that the approximation is introduced. This procedure allows to obtain an analytical energy formula which reads 
\begin{equation}
E^H(\rho)=\sqrt{\frac{2\rho}{m}}\, (2n+\ell+3/2)+ \frac{2-\lambda}{2\lambda} 
(2 \rho)^{\frac{\lambda}{\lambda-2}}(a|\lambda|)^{\frac{2}{2-\lambda}}.
\end{equation}
The wave function is also readily known from~(\ref{nrho1}) in which $\omega$ is replaced by $\sqrt{2\rho/m}$. We will not write explicitly the wave functions in the following, since they can readily be obtained once the energy is known. 

The last step is the minimization of the energy with respect to the auxiliary field $\rho$. One finds 
\begin{equation}
\label{r0def2}
\rho_0=\frac{1}{2}\left[\frac{2n+\ell+3/2}{\sqrt{m}} \right]^{\frac{2(\lambda-2)}{\lambda+2}} (a|\lambda|)^{\frac{4}{\lambda+2}},
\end{equation}
and then the final form for the energy spectrum is 
\begin{equation}
\label{eh}	
E^H(\rho_0)=\left(\frac{2+\lambda}{2\lambda}\right)\frac{(a\, |\lambda|)^{\frac{2}{\lambda+2}}}{m^{\frac{\lambda}{\lambda+2}}}
(2n+\ell+3/2)^{\frac{2\lambda}{\lambda+2}}.
\end{equation}
It is interesting to notice that 
\begin{eqnarray}
\left.E^H(\rho_0)\right|_{\lambda=2}&=&\sqrt{\frac{2a}{m}}(2n+\ell+3/2),\\
\left.E^H(\rho_0)\right|_{\lambda=-1}&=&-\frac{ma^2}{2(2n+\ell+3/2)^2}.
\end{eqnarray}
The formula is exact for $\lambda=2$ as expected, since in this case $V(r)=P(r)$. If, in the case $\lambda=-1$, we substitute $2n+3/2$ by $n+1$, we recover the exact result for the Coulomb case. Of course, the presence of a term $2n+l+3/2$ is a ``remembrance'' of the original function $P(r)$ that we employed to treat the auxiliary field. The fact that a potential with a very strong increase for large $r$ can mimic the asymptotic behavior of the energies for a potential that tends to zero for large $r$ is astonishing and shows the power of our method to get at least the good asymptotic properties. We remark also that formula~(\ref{eh}) is not defined when $\lambda=-2$. Moreover, the energy becomes positive when $\lambda < -2$. We recover actually a well-known result: In an attractive potential with a singularity like $1/r^n$ with $n\ge 2$, the existence of bound states is not automatically guaranteed \cite{case50}.  

We mentioned in the previous section that $\rho_0$ was expected to be close to the average value of $\hat\rho$. The average should here be computed with the wave functions $R^H(r)$ given by~(\ref{nrho1}) in which $\omega=\sqrt{2\rho_0/m}$. This point can be explicitly checked in the particular case we are dealing with. Indeed, relation~(\ref{rdpl}) shows that 
\begin{equation}
\left\langle \hat\rho\right\rangle=\frac{a|\lambda|}{2} \left\langle  r^{\lambda-2}\right\rangle\approx\frac{a|\lambda|}{2} \left\langle r^2\right\rangle^{\frac{\lambda-2}{2}} .
\end{equation}
But, it is well-known that for the harmonic oscillator~(\ref{nrho1}), one has $\left\langle r^2\right\rangle=(2n+\ell+3/2)/(m\omega)$. Thus in our case, $\left\langle r^2\right\rangle=(2n+\ell+3/2)/\sqrt{2m\rho_0}$. Using~(\ref{r0def2}), we find after some algebra that
\begin{equation}
\left\langle \hat\rho\right\rangle\approx \rho_0.
\end{equation}
This confirms the interpretation of the auxiliary field technique as a mean field approximation in which the corresponding field is chosen so that the calculations are as easy as possible.

Using the results developed above, applied to (\ref{errana2}), one can compute that, for the ground state,
\begin{equation}
\label{ineq1}
\frac{E(\rho_0)-E}{|E(\rho_0)|} \ge
\textrm{sgn}(\lambda)\frac{2}{\lambda+2}\left[ 1- \frac{2}{\sqrt{\pi}}\Gamma\left( \frac{\lambda+3}{2} \right)
\left( \frac{2}{3} \right)^{\lambda/2}\right].
\end{equation}
For $\lambda=2$, the trivial equality $0=0$ is found since the method is exact in this case. For $\lambda=1$, it has been numerically verified that the l.h.s.\ and r.h.s.\ of this equation are respectively $0.0559$ and $0.0525$. For $\lambda=-1$, these numbers become respectively $1.25$ and $0.764$. In both cases, the formula is verified. As expected, the relative error is smaller for $\lambda=1$ than for $\lambda=-1$, since the harmonic potential is ``closer" to potential $r$ than potential $-1/r$. Since $E(\rho_0)$ and $E$ obey the same scaling laws (see section~\ref{scale}), inequality~(\ref{ineq1}) does not depend on parameters $m$ and $a$.

\subsection{Logarithmic potential}

We turn now our attention to the special case of a Schr\"{o}dinger equation with the logarithmic potential ($b > 0$)
\begin{equation}
\label{plog}
V(r)=a\, \ln(b\,r).
\end{equation}
With the choice $P(r)=r^2$, equation~(\ref{rdef}) can be easily solved. One finds
\begin{equation}
r = K^{-1}(\hat\rho)=\sqrt{\frac{a}{2\hat\rho}}
\end{equation}
and ($e$ is the basis of natural logarithms)
\begin{equation}
E(\rho)=\sqrt{\frac{2\rho}{m}}\, (2n+\ell+3/2)+\frac{a}{2} \ln\left[\frac{a\, b^2}{2\, e\, \rho}\right].	
\end{equation}
The elimination of $\rho$ from the minimization condition of this last energy formula finally leads to
\begin{eqnarray}
\rho_0&=&\frac{m\, a^2}{2(2n+\ell+3/2)^2}, \\
\label{elog}
E(\rho_0)&=&a\, \ln\left[\sqrt{\frac{e}{m\,a}}\, b\, (2n+\ell+3/2)\right].
\end{eqnarray}

It is well known that the eigenenergies $E(m,a,b;n,l)$ resulting from a Schr\"{o}dinger equation with the potential~(\ref{plog}) satisfy the property \cite{QR}
\begin{equation}
\label{proplogpot}
E(\alpha m,a,b;n,l)=E(m,a,b;n,l) - \frac{a}{2} \ln \alpha.
\end{equation}
The immediate consequence is that the corresponding spectrum is independent of the mass of the particle. It is remarkable that the basic property~(\ref{proplogpot}) still holds for our approximate expression~(\ref{elog}).

It is worth mentioning that the energy formula~(\ref{elog}) can be understood as a particular limit case of formula~(\ref{eh}), as it was suggested in~\cite{lucha}. The basic idea is that the logarithmic potential can be rewritten into a power-law form, i.e. 
\begin{equation}
\ln\, x=\lim_{\lambda\rightarrow0}\frac{1}{\lambda}(x^\lambda-1).
\end{equation}
Let us then consider the following Hamiltonian
\begin{equation}
\label{hlog2}
H(\lambda)=\frac{\bm p^{\, 2}}{2m}+\frac{a}{\lambda}\left[(b\, x)^{\lambda}-1\right].
\end{equation}
The potential term of this Hamiltonian reduces to potential~(\ref{plog}) in the limit $\lambda\rightarrow 0$. A simple rewriting of formula~(\ref{eh}) for $a \rightarrow a\, b^\lambda/|\lambda|$ gives the eigenenergies of Hamiltonian~(\ref{hlog2}). We have thus
\begin{equation}
E(\lambda)=\left(\frac{2+\lambda}{2\lambda}\right) \frac{(a\, b^\lambda)^{\frac{2}{\lambda+2}}}{m^{\frac{\lambda}{\lambda+2}}} (2n+\ell+3/2)^{\frac{2\lambda}{\lambda+2}}-\frac{a}{\lambda},
\end{equation}
and as expected
\begin{equation}
\label{liml0}
\lim_{\lambda\rightarrow0}E(\lambda)=a\, \ln\left[\sqrt{\frac{e}{m\,a}}\, b\, (2n+\ell+3/2)\right],
\end{equation}
that is precisely formula~(\ref{elog}), directly obtained from the logarithmic potential. This confirms the idea that the logarithmic potential can be seen as the limit of a power-law potential $r^\lambda$ when $\lambda$ goes to zero.    

\section{The case $P(r)=-1/r$}\label{capp}

\subsection{Power-law potentials}

Another obvious case of analytically solvable radial potential is the Coulomb potential
\begin{equation}\label{coul}
V^{C}(r)=-\frac{\kappa}{r},
\end{equation}
whose eigenfunctions and eigenenergies respectively read (see for example~\cite[problem 67]{flu})
\begin{equation}
\label{nrc}
R^C(r)=N_{n\ell}\, r^\ell\ e^{-\gamma r}\  L^{2\ell+1}_{n}(2 \gamma r),\quad E^{C}=-\frac{m\kappa^2}{2(n+\ell+1)^2},
\end{equation}
with $\gamma=m\kappa/(n+\ell+1)$. 

It is thus of interest to choose $P(r)=-1/r$ and to perform the same calculations as the ones we did in the previous section. Consequently, we will give less details and only mention the main steps of the computations. Equation~(\ref{rdef}) takes the form
\begin{equation}\label{rdpl2}
\hat\rho=a|\lambda|\, r^{\lambda+1}\Rightarrow r = K^{-1}(\hat\rho)=\left(\frac{\hat\rho}{a|\lambda|}\right)^{\frac{1}{\lambda+1}},
\end{equation}
and the energy spectrum is analytically known once the auxiliary field $\rho$ is introduced as in~(\ref{htdef})
\begin{eqnarray}\label{r0coul}
\rho_0&=&\left[\frac{1}{m}\, \left(a|\lambda|\right)^{\frac{1}{\lambda+1}}(n+\ell+1)^2\right]^{\frac{\lambda+1}{\lambda+2}}, \\
\label{ec}
E^C(\rho_0)&=&\left(\frac{2+\lambda}{2\lambda}\right) \frac{(a\, |\lambda|)^{\frac{2}{\lambda+2}}}{m^{\frac{\lambda}{\lambda+2}}} (n+\ell+1)^{\frac{2\lambda}{\lambda+2}}.
\end{eqnarray}
Surprisingly, formula~(\ref{ec}) is identical to~(\ref{eh}), but the $(n+\ell+1)$ factor due to the use of the Coulomb potential $P(r)$ is now present. The $(2n+\ell+3/2)$ or $(n+\ell+1)$ factor is thus a ``remembrance" of the harmonic or Coulomb approximation. However, the same qualitative behaviors in $\ell^{-2}$ and $n^{-2}$ are obtained. In particular,
\begin{eqnarray}
\left.E^C(\rho_0)\right|_{\lambda=2}&=&\sqrt{\frac{2a}{m}}(n+\ell+1), \\
\left.E^C(\rho_0)\right|_{\lambda=-1}&=&-\frac{ma^2}{2(n+\ell+1)^2}.
\end{eqnarray}
As expected, this method is exact for the Coulomb potential.  If, in the formula for $\lambda=2$, we substitute $n+1$ by $2n+3/2$ we recover the exact result. Once more, it is remarkable that using a potential $P(r)$ which decreases at large $r$, one can mimic the behavior of the eigenenergies due to a potential $V(r)$ which strongly increases at large $r$. It is worth mentioning that, for large $n$ and $\ell$ values, both $E^H(\rho_0)$ and $E^C(\rho_0)$ are proportional to $n^{2\lambda/(\lambda+2)}$ and $\ell^{2\lambda/(\lambda+2)}$. Such an asymptotic behavior is in agreement with the WKB analysis which is performed in~\cite{QR}. The same behavior is also found in \cite{bose}.

The interpretation of the auxiliary field method as a particular mean field approximation would be strengthened if we had $\left\langle \hat\rho\right\rangle\approx\rho_0$ as in the previous section, with $\hat\rho$ and $\rho_0$ given by~(\ref{rdpl2}) and (\ref{r0coul}) respectively. The average value should actually be computed with the eigenstates of Hamiltonian~(\ref{htdef}) in the particular case we are dealing with. These eigenstates are actually eigenstates of the following Hamiltonian
\begin{equation}
H_A=\frac{\bm p^{\, 2}}{2m}-\frac{\rho_0}{r},
\end{equation}
whose eigenenergies read
\begin{equation}
E_A=-\frac{m\, \rho^2_0}{2(n+\ell+1)^2}.
\end{equation}
By applying the virial theorem to this last Hamiltonian, one obtains $E_A=-(\rho_0/2)\left\langle1/r \right\rangle\approx -\rho_0/(2\left\langle r\right\rangle)$. Thus, one has
\begin{equation}
\left\langle r\right\rangle\approx\frac{(n+\ell+1)^2}{m\rho_0},
\end{equation}
 and 
\begin{equation}
\left\langle \hat\rho\right\rangle=a\, |\lambda|\, \left\langle r^{\lambda+1}\right\rangle\approx a\, |\lambda|\, \left\langle r\right\rangle^{\lambda+1}.
\end{equation}
A straightforward calculation shows that these two relations together with the definition~(\ref{r0coul}) leads to $\left\langle \hat \rho\right\rangle\approx \rho_0$, as expected. 

Using the results developed above, applied to (\ref{errana2}), one can compute that, for the ground state,
\begin{equation}
\label{ineq2}
\frac{E(\rho_0)-E}{|E(\rho_0)|} \ge
\textrm{sgn}(\lambda)\frac{2}{\lambda+2}\left[ 1- \frac{\Gamma(\lambda+3)}{2^{\lambda+1}} \right].
\end{equation}
For $\lambda=-1$, the trivial equality $0=0$ is found since the method is exact in this case. For $\lambda=1$, it has been numerically verified that the l.h.s.\ and r.h.s.\ of this equation are respectively $-0.237$ and $-0.333$. For $\lambda=2$, these numbers become respectively $-1/2$ and $-1$. In both cases, the formula is verified. As expected, the relative error is smaller for $\lambda=1$ than for $\lambda=2$, since the Coulomb potential is ``closer" to potential $r$ than potential $r^2$. Since $E(\rho_0)$ and $E$ obey the same scaling laws (see section~\ref{scale}), inequality~(\ref{ineq2}) does not depend on parameters $m$ and $a$.

\subsection{Logarithmic potential}

The resolution of the Schr\"{o}dinger equation with the logarithmic potential~(\ref{plog}) is even simpler with the present choice of $P(r)$. Indeed, one finds in this case
\begin{equation}
r = K^{-1}(\hat\rho)= \frac{\hat\rho}{a},
\end{equation}
and 
\begin{equation}
E(\rho)=-\frac{m \rho^2}{2(n+\ell+1)^2}+a \ln \left( \frac{b\, \rho\, e}{a} \right).	
\end{equation}
The elimination of $\rho$ finally leads to 
\begin{eqnarray} 
\rho^2_0&=&\frac{a (n+\ell+1)^2}{m}, \\
\label{elog2}
E(\rho_0)&=&a\, \ln\left[\sqrt{\frac{e}{m\,a}}\, b\, (n+\ell+1)\right].
\end{eqnarray}
Again, formulae~(\ref{elog}) and (\ref{elog2}) exhibit the same formal expression but a $(n+\ell+1)$ factor is now present. Consequently, the properties mentioned above concerning the spectrum and the limit of a power-law potential when $\lambda \to 0$ still hold in this case.

\section{Scaling laws}
\label{scale}

Let us consider the Schr\"{o}dinger equation
\begin{equation}
\label{sl1}
\left[\frac{\bm p^{\, 2}}{2m}+\gamma\, V(\alpha\, r)-E(m,\gamma,\alpha)\right]\, \Psi(\bm r\,)=0,
\end{equation}
where $\gamma$ and $\alpha$ are two real parameters. This last equation can be rewritten with a different set of parameters (denoted as $m'$, $\gamma'$, and $\alpha'$) and with the change of variables $\bm r=\beta\, \bm q$. After a further multiplication by the constant $\chi$, we obtain
\begin{equation}
\label{sl2}
\left[\chi\frac{\bm p^{\, 2}_q}{2m'\beta^2}+\chi\gamma'\, V(\alpha'\, \beta\, q)-\chi E(m',\gamma',\alpha')\right]\, \tilde\Psi(\bm q\,)=0.
\end{equation} 

It is readily checked that, with $\beta=\alpha/\alpha'$ and $\chi=\beta^2 m'/ m$,~(\ref{sl2}) reduces to~(\ref{sl1}) provided that the following scaling laws are assumed
\begin{eqnarray}
\label{sl3}
\gamma&=&\gamma'\, \left(\frac{\alpha}{\alpha'}\right)^2\, \frac{ m'}{ m}, \\
\label{sl4}
E(m,\gamma,\alpha)&=&\left(\frac{\alpha}{\alpha'}\right)^2\, \frac{ m'}{ m}\, E( m',\gamma',\alpha').
\end{eqnarray}
This is the expression of the scaling law for the most general potential $V(r)$. However, in the special case of power-law potentials $V(r)=a r^\lambda$, we have a more powerful expression due to the fact that we have some freedom on $\gamma$ and $\alpha$ parameters. Indeed, if these parameters are chosen with the constraint $\gamma \alpha^\lambda=a$ they lead to the same potential. This property allows to obtain the absolute dependence of the energies in term of $m$ and $a$ parameters. Explicitly, we can write
\begin{equation}
\label{scl1}
E(m,a;\lambda,n,\ell)= 2^\frac{\lambda}{\lambda+2} 
a^{\frac{2}{\lambda+2}} m^{-\frac{\lambda}{\lambda+2}}\, \epsilon(\lambda,n,\ell),
\end{equation}
where $\epsilon(\lambda,n,\ell)$ is an eigenvalue of the dimensionless Schr\"{o}dinger equation
\begin{equation}
\label{ham_dim}
H=\frac{\bm q\,^2}{4}+ \textrm{sgn}(\lambda) \, |\bm x\,|^\lambda. 
\end{equation}
This Hamiltonian corresponds to the particular values $m=2$ and $a=1$. This choice is made such that, in the exact cases, $\epsilon(2,n,\ell)=2n+\ell+3/2$ and $\epsilon(-1,n,\ell)=-(n+\ell+1)^{-2}$.
One sees that equations~(\ref{eh}) and (\ref{ec}) resulting from our treatment fulfills exactly this scaling law~(\ref{scl1}), and that the dimensioned factor gives the correct dimension to the energy $E(m,a;\lambda,n,\ell)$. 

So an approximate expression for $\epsilon(\lambda,n,\ell)$, that is deduced from the auxiliary field technique, can be written as
\begin{equation}
\label{egsolap}
\epsilon^{\textrm{(app)}}(\lambda,n,\ell)= \left(\frac{2+\lambda}{2\lambda}\right) 
\frac{|\lambda|^{\frac{2}{\lambda+2}}}{2^{\frac{\lambda}{\lambda+2}}}
\left( N^{\textrm{(app)}}(n,\ell) \right)^{\frac{2 \lambda}{\lambda+2}},
\end{equation}
with the values $N^{(H)}(n,\ell)=2n+l+3/2$ for the approximation based on the harmonic function $P(r)=r^2$ and $N^{(C)}(n,\ell)=n+l+1$ for the approximation based on the Coulomb function $P(r)=-1/r$. In the next section we will show how to improve the results suggested by the previous studies.
 
\section{Improved formulae}\label{impfor}

Following the form chosen for $P(r)$, two different approximations (\ref{egsolap}) are obtained for both power-law and logarithmic potentials. A comparison with exact results (numerically computed) shows that these formulae are satisfactory but it is possible to improve the results. Using scaling laws presented above, it is sufficient to focus our attention on the $\epsilon^{\textrm{(app)}}(\lambda,n,\ell)$ quantity.

We propose here the following expression
\begin{equation}
\label{form1}
\epsilon^{(A)}(\lambda,n,\ell)= \left(\frac{2+\lambda}{2\lambda}\right) 
\frac{|\lambda|^{\frac{2}{\lambda+2}}}{2^{\frac{\lambda}{\lambda+2}}}
\left[ b(\lambda)\,n + \ell + c(\lambda)
\right]^{\frac{2 \lambda}{\lambda+2}}.
\end{equation}
Since this formula is exact for $\lambda=-1$ and $\lambda=2$, we must have $b(-1)=1$, $c(-1)=1$, $b(2)=2$, and $c(2)=3/2$. Moreover, we can expect that the use of $P(r)=-1/r$ will give better results for negative values of $\lambda$, while the use of $P(r)=r^2$ will give better results for positive values of $\lambda$. So a natural choice would be 
\begin{eqnarray}
\label{bc1}
b(\lambda)= 2, \quad c(\lambda) =3/2, & \quad \textrm{for}& \quad \lambda > 0, \nonumber \\
b(\lambda)= 1, \quad c(\lambda) =1,   & \quad \textrm{for}& \quad -2 < \lambda < 0 .
\end{eqnarray}
The variations of the energy being smooth for variations of $\lambda$, we can refine our assumptions and suppose that $b(\lambda)$ and $c(\lambda)$ are continuous functions of this parameter (the particular case $\lambda=0$ will be discussed with the logarithmic potential). So one can first think to a linear interpolation between the two analytical cases 
\begin{equation}
\label{bc2}
b(\lambda)= \frac{\lambda+4}{3}, \quad 
c(\lambda)= \frac{\lambda+7}{6} .
\end{equation}

\begin{figure}[ht]
\begin{center}
\includegraphics*[width=10cm]{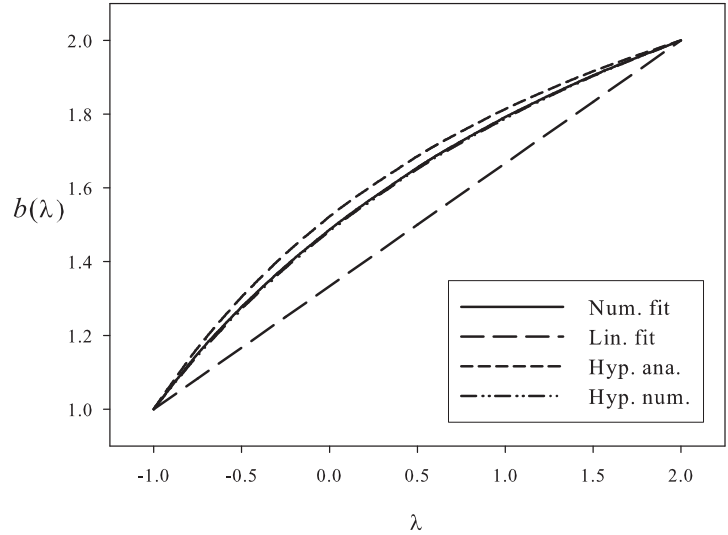}
\caption{Coefficient $b(\lambda)$ from~(\ref{form1}) as a function of $\lambda$, obtained from various approximations: numerical fit minimizing the function $\chi(\lambda)$ (solid line), linear fit from~(\ref{bc2}) (long-dashed line), hyperbola from~(\ref{bc3}) (short-dashed line), hyperbola from~(\ref{bc4}) (dashed-dotted line). This last curve is nearly indistinguishable from the solid line. }
\label{fig1}
\end{center}
\end{figure}

\begin{figure}[ht]
\begin{center}
\includegraphics*[width=10cm]{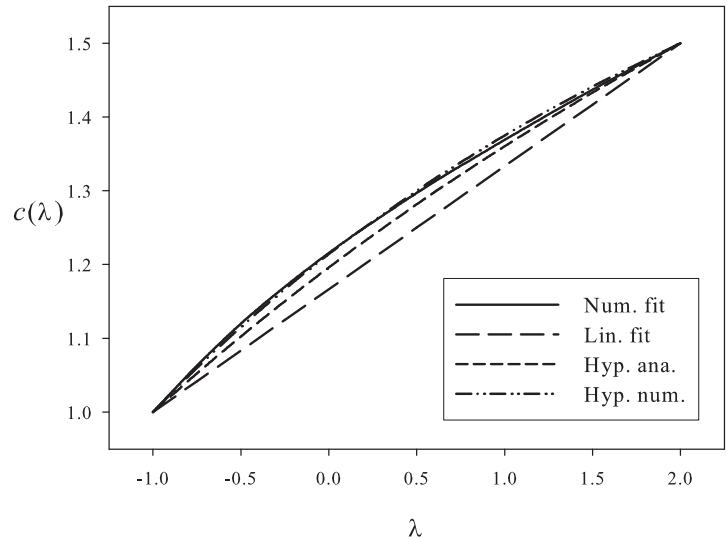}
\caption{Same as figure~\ref{fig1}, but for the coefficient $c(\lambda)$ from~(\ref{form1}).}
\label{fig2}
\end{center}
\end{figure}

Nevertheless, it is possible to greatly improve the accuracy of the approximation~(\ref{form1}) by looking at  numerical results. Very accurate eigenvalues $\epsilon_{\textrm{num}}(\lambda,n,\ell)$ of Hamiltonian~(\ref{ham_dim}) can be obtained numerically with the Lagrange mesh method \cite{lag}. In order to find the best possible values for the coefficients $b(\lambda)$ and $c(\lambda)$, we will use the following measure
\begin{equation}
\label{chi2}
\chi(\lambda)=\sum_{n=0}^{3}\sum_{\ell=0}^{3} \left( \epsilon_{\textrm{num}}
(\lambda,n,\ell) - \epsilon^{(A)}(\lambda,n,\ell) \right)^2.
\end{equation}
Other choices are possible but we find this one very convenient. The coefficients $b(\lambda)$ and $c(\lambda)$ minimizing the quantity $\chi(\lambda)$ have been computed for several values of $\lambda$ (see figures~\ref{fig1} and \ref{fig2}). In both cases, the curve is similar to a section of conic. 

To determine analytically the best conics fitting the coefficients, it is necessary to obtain a third known point on each curve. Solutions of the case $\lambda=1$ are known for $\ell=0$ in terms of zeros of the Airy function. Approximate eigenvalues of the Hamiltonian~(\ref{ham_dim}) with a linear potential are given by \cite{sema04,abra70} (this is an asymptotic form valid for large $n$)
\begin{equation}
\label{airy1}
\epsilon_{\textrm{Ai}}(1,n,0)=\left( \frac{3\pi}{4} \right)^{2/3}\left( n +\frac{3}{4} \right)^{2/3}.
\end{equation}
Equating relations~(\ref{airy1}) and (\ref{form1}), we find $b(1)=\pi/\sqrt{3}$ and $c(1)=\sqrt{3}\pi/4$. Assuming that this procedure can be extended for any values of $\ell$, the conics can be determined by using the exact values $b(-1)=1$, $c(-1)=1$, $b(2)=2$, and $c(2)=3/2$. We find that hyperbolae give the best results (minimal value of $\chi(\lambda)$). The coefficients are then given by 
\begin{eqnarray}
\label{bc3}
\sigma&=&\sqrt{3}\pi , \nonumber \\
b(\lambda)&=& \frac{(4\sigma-18) \lambda+(18-2\sigma)}
{(3\sigma-15) \lambda+(21-3\sigma)}, \nonumber \\
c(\lambda)&=& \frac{(7\sigma-36) \lambda+(36-5\sigma)}{(6\sigma-32) \lambda+(40-6\sigma)}.
\end{eqnarray}
These relations could look quite complicated but it is worth noting the symmetries existing between the absolute values of the numbers present in the formula. When a number appears twice in a coefficient of the numerator (denominator), it is the arithmetic mean of two other corresponding numbers in the denominator (numerator). 
A number which appears twice in $c(\lambda)$ is the double that the corresponding number in $b(\lambda)$. The quality of this approximation can be estimated in figures~\ref{fig1} and \ref{fig2}, and in table~\ref{tab1} where results obtained from the following formulae (computed in the framework of the WKB method) are also given~\cite{QR}
\begin{eqnarray}
\label{WKB}
\epsilon_{\textrm{WKB}}(\lambda,n,\ell)&=&+\left( \frac{\lambda\, \sqrt{\pi}\, \Gamma\left( \frac{3}{2}+\frac{1}{\lambda} \right)}{2\, \Gamma\left( \frac{1}{\lambda} \right)} \left( 2\, n + \ell + \frac{3}{2} \right) \right)^{\frac{2 \lambda}{\lambda+2}} \quad \textrm{for} \quad \lambda > 0, \nonumber \\
&=&-\left( \frac{|\lambda|\, \sqrt{\pi}\, \Gamma\left( 1-\frac{1}{\lambda} \right)}{2\, \Gamma\left( -\frac{1}{2} -\frac{1}{\lambda} \right)} \left( 2\, n + 2 - \frac{1+\lambda-2\,\ell}{2+\lambda} \right) \right)^{\frac{2 \lambda}{\lambda+2}}\nonumber\\
&&\qquad \textrm{for} \quad -2 < \lambda < 0 .
\end{eqnarray}

It is clear that the approximations are mainly good for $-1 \le \lambda \le 2$. 
Let us note that $\epsilon_{\textrm{WKB}}(1,n,0)=\epsilon_{\textrm{Ai}}(1,n,0)$. Thus, approximations~(\ref{WKB}) and (\ref{form1}) with (\ref{bc3}) give the same results in the cases $\lambda=2$, $-1$ (exact results), and $\lambda=1$ for $\ell=0$ (approximations that tend to the exact results for large $n$); however they differ for all other sets ($\lambda,n,\ell$). Our formulation has the advantage to deal with a unique consistent formula, valid for any $\lambda$, whereas the WKB approximation requires two different expressions, one for $\lambda>0$ and one for $\lambda<0$. A comparison with exact results (see table~\ref{tab1} and table~\ref{tab2}) shows that our formula is better than the WKB formula in any case.

Other approximate forms for the eigenvalues of Hamiltonian~(\ref{ham_dim}) exist and give good results \cite{bose,fabre}, but they are not considered here because they do not give the exact result for the cases $\lambda=-1$ and $\lambda=2$.

\begin{table}[ht]
\begin{center}
\caption{Values of $\chi(\lambda)$ defined by~(\ref{chi2}), as a function of $\lambda$, for various approximations. $\chi(-1)=\chi(2)=0$ in all cases. The value $\lambda=0$ corresponds to the logarithmic potential (see text).}
\label{tab1}
\begin{tabular}{cccc}
\hline
$\lambda$ & \multicolumn{3}{c}{$\chi(\lambda)$} \\
          & Equations~(\ref{form1}) with (\ref{bc3}) & Equations~(\ref{form1}) with (\ref{bc4}) & Equation~(\ref{WKB}) \\
\hline
4 & 0.44 & 0.23 & 0.88 \\
3/2 & 4.0~10$^{-3}$ & 1.7~10$^{-3}$ & 3.1~10$^{-2}$ \\
1 & 5.1~10$^{-3}$ & 2.1~10$^{-3}$ & 6.9~10$^{-2}$ \\ 
2/3 & 2.8~10$^{-3}$ & 1.1~10$^{-3}$ & 5.9~10$^{-2}$ \\ 
1/2 & 1.5~10$^{-3}$ & 5.7~10$^{-4}$ & 4.2~10$^{-2}$ \\
1/10 & 3.3~10$^{-5}$ & 1.1~10$^{-5}$ & 2.0~10$^{-3}$ \\
0 & 2.1~10$^{-3}$ & 7.2~10$^{-4}$ & - \\ 
$-1/4$ & 4.9~10$^{-5}$ & 1.5~10$^{-5}$ & 2.6~10$^{-3}$ \\
$-1/2$ & 9.2~10$^{-5}$ & 2.2~10$^{-5}$ & 2.6~10$^{-3}$ \\
$-3/4$ & 1.0~10$^{-4}$ & 1.7~10$^{-5}$ & 1.0~10$^{-3}$ \\
$-3/2$ & 27 & 8 & 30 \\
\hline
\end{tabular}
\end{center}
\end{table}

It is possible to improve the value of the quantity $\chi(\lambda)$ by fitting the hyperbolae $b(\lambda)$ and $c(\lambda)$ directly with the numerical values $\epsilon_{\textrm{num}}(\lambda,n,\ell)$. Several choices are possibles. We give here a parameterization with simple numbers (below 100) which give good results 
\begin{equation}
\label{bc4}
b(\lambda)= \frac{41 \lambda+86}{13 \lambda+58}, \quad 
c(\lambda)= \frac{5 \lambda+17}{2 \lambda+14} .
\end{equation}
The quality of this fit can be appraised in figures~\ref{fig1} and \ref{fig2}, and in table~\ref{tab1}. Obviously the numbers in~(\ref{bc4}) depend on the points used for the fit but also on the particular choice of the function $\chi(\lambda)$. Other definitions, relative error instead of absolute error or different summations on quantum numbers, would have given other numbers. Parameterization~(\ref{bc4}) gives better global results than parameterization~(\ref{bc3}), except for small values of $\ell$ and large values of $n$. This behavior is expected since formulae~(\ref{bc3}) rely on the formula~(\ref{airy1}) whose accuracy increases with $n$ \cite{abra70}.

In order to have a better estimate of the qualities of our results, we compare in table~\ref{tab2} the dimensionless energies $\epsilon(\lambda=1,n,\ell)$ obtained from various approximations. One can remark that equation~(\ref{form1}) with approximation~(\ref{bc3}) gives good results, as good as or better than $\epsilon_{\textrm{WKB}}$, and that equation~(\ref{form1}) with approximation~(\ref{bc4}) gives very good overall results, much better than the previous ones for $\ell \neq 0$. This is particularly significant for large values of $n$ and $\ell$ quantum numbers. These conclusions remain valid for any value of $\lambda$.

\begin{table}[ht]
\begin{center}
\caption{Values of $\epsilon(\lambda=1,n,\ell)$ for various approximations. For each set $(n,\ell)$, 
first line is the result obtained by numerical integration [$\epsilon_{\textrm{num}}(1,n,\ell)$], 
second line by equations~(\ref{form1}) and (\ref{bc3}), 
third line by equations~(\ref{form1}) and (\ref{bc4}), 
and fourth line by equation~(\ref{WKB}) [$\epsilon_{\textrm{WKB}}(1,n,\ell)$]. }
\label{tab2}

\begin{tabular}{ccccc}
\hline
 $\ell$ & $\epsilon(1,0,\ell)$ & $\epsilon(1,1,\ell)$ & $\epsilon(1,2,\ell)$ & $\epsilon(1,3,\ell)$ \\
\hline
 0 & 1.47292 &  2.57525 &  3.47773 &  4.27536 \\
   & 1.46167 &  2.57138 &  3.47560 &  4.27394 \\
   & 1.47214 &  2.56575 &  3.45909 &  4.24854 \\
   & 1.46167 &  2.57138 &  3.47560 &  4.27394 \\
\hline
 1 & 2.11746 &  3.07701 &  3.91056 &  4.66528 \\
   & 2.11057 &  3.08645 &  3.92585 &  4.68320 \\
   & 2.11929 &  3.08132 &  3.91032 &  4.65894 \\
   & 2.05470 &  3.04039 &  3.88505 &  4.64587 \\
\hline
 2 & 2.67619 &  3.54649 &  4.32712 &  5.04580 \\
   & 2.67098 &  3.56156 &  4.35159 &  5.07529 \\
   & 2.67874 &  3.55678 &  4.33684 &  5.05198 \\
   & 2.57138 &  3.47560 &  4.27394 &  5.00345 \\
\hline
 3 & 3.18188 &  3.98898 &  4.72763 &  5.41584 \\
   & 3.17757 &  4.00682 &  4.75742 &  5.45277 \\
   & 3.18469 &  4.00231 &  4.74332 &  5.43029 \\
   & 3.04039 &  3.88505 &  4.64587 &  5.34868 \\
  \hline 
\end{tabular}
\end{center}

\end{table}

Let us now consider the dimensionless logarithmic potential
\begin{equation}
\label{ham_dim_ln}
H=\frac{\bm q\,^2}{4}+ \ln |\bm x\,|. 
\end{equation}
Using the results above as well as~(\ref{liml0}), it is clear that a good approximate form for the eigenvalues is 
\begin{equation}
\label{form2}
\epsilon(n,\ell)=\ln \left[ \sqrt{\frac{e}{2}}\left( b(0)\,n + \ell + c(0) \right) \right].
\end{equation}
The quality of this formula can be estimated in table~\ref{tab1}. 

\section{Conclusions}
\label{conclu}

In this paper, we propose a new method to obtain approximate expressions for the eigenenergies of a Schr\"{o}dinger equation based on any local central potential $V(r)$ and valid for any radial $n$ and orbital $\ell$ quantum numbers. The method relies on the introduction of an auxiliary field $\rho$. Given a potential $P(r)$ for which the analytical spectrum of the Schr\"{o}dinger equation is known, we presented an algorithm which transforms the solvable Hamiltonian into the Hamiltonian which contains the original potential $V(r)$ and for which the eigenenergies are not known analytically. This transformation needs an auxiliary field which makes a bridge between $V(r)$ and $P(r)$. If the auxiliary field is considered as an operator $\hat{\rho}$ which cancels the variation of the transformed Hamiltonian, the algorithm gives the exact solution. We propose to start from the solvable Hamiltonian and the corresponding eigenenergies expressed in terms of the auxiliary field, and consider this field not as an operator but as a pure number. If this number is chosen by a minimization procedure, the resulting energies gives a good approximation to the exact values. If $P(r)=V(r)$ our method gives the exact results.

This method is applied to a power-law type of potential with power $\lambda$. The resulting expression for the energies satisfies the scaling law prescriptions and gives the good asymptotic behavior for the radial and orbital quantum numbers. Comparing the results coming from the use of $P(r)=r^2$ and $P(r)=-1/r$, one sees that they are identical provided that the $(2n+l+3/2)$ term, remembrance of the harmonic oscillator, is replaced by $(n+l+1)$, remembrance of the Coulomb potential. Owing to this remarkable property, we proposed new analytical formulae, valid for any $\lambda$, $n$, $\ell$, that give results much better than the equivalent ones which could be found in literature since they are able to reproduce the exact results with an accuracy often better than $10^{-3}$. These formulae can be used for example if one wishes to determine the parameters of a power-law potential that fits the experimental data as better as possible. We also showed that the results for logarithmic potentials are obtained very easily from those of the power-law potential if one takes the limit $\lambda \to 0$.

The main interest of our method is to obtain analytical formulae. Its main difficulty consists of the inversion of some algebraical equations. In the WKB method, analytical formulae can also be obtained provided the integral over the radial momentum can be performed. Depending on the particular problem considered, one or the other method can be more efficient. So the new method based on an auxiliary field, proposed here, can be a useful supplementary technique to solve eigenvalue problems. It seems very promising to obtain analytical expressions for the eigenenergies of a Schr\"{o}dinger equation with a number of interesting potentials. Some of them are under study. Analytical approximate eigenvectors are also given by this method. Their study will be considered in a subsequent paper. 

\section*{Acknowledgments}
C. S. and F. B. would thank the F.R.S.-FNRS for financial support. C. S. would thank F. Brau pour useful discussions. The authors would also thank an anonymous referee for a useful suggestion.

\end{document}